\documentclass{amsart}
\usepackage[utf8]{inputenc}
\usepackage{graphicx}

\title{Smart transport infrastructure maintenance. 
\\A smart-contract blockchain approach. }
\author{Fatjon Seraj\\University of Twente \\f.seraj@utwente.nl}

\date{October 2018}

\begin{document}

\maketitle
\begin{abstract}
     Infrastructure maintenance is an intrinsically complicated procedure. The transport infrastructure such the roads and railroads are complex and widely dispersed infrastructures. Maintaining this infrastructure involves many partners working in concert to provide a safe and efficient infrastructure, maintained according to technical and safety standards, with suitable materials provided on time and within budget. Traditionally these requirements are sealed in the paper, and each contract step is checked manually or physically. 
     Smart contracts are the new iteration of contracts based on the blockchain distributed ledger technology. Distributed ledgers facilitate secure and transparent transactions between parties in a decentralized fashion, where if the condition is met, the parties conclude the transaction. When applied beyond the primary financial transaction, blockchains allow more complex agreements, such as keeping track of the fulfillment of a contract between two or more parties.
     smart contract is a set of rules written in computer code stored in the blockchain, and the execution of each of the contract terms is auto-executed when the condition is met and recorded on the blockchain. This way smart contracts facilitate and automate the whole infrastructure maintenance process, form the moment a contractor is assigned to maintain, improve or build the segment up to the accomplishment of the maintenance term. Depending upon a decentralized, immutable record, the terms and status of the smart contract is equally available to all the parties involved in the agreement assuring a higher level of trust and efficiency. 
     Compiling a smart contract for infrastructure maintenance necessitates a deep understanding of the whole procedural chain involved in the maintenance so we can foresee all the requirements and liabilities. This chain of procedures includes smart continuous monitoring of the infrastructure through a data-driven dynamic maintenance model, which will automatically trigger the maintenance of the segment in need. Modern process mining techniques can create an adaptive, robust, and generic Maintenance Process Model that can help the Operations Management compile the necessary contract rules such  as allocating the required assets, logistics, required materials to be used, and the needed skills to handle the procedure. To automatically inspect and certify these steps, smart contracts require the involvement of modern technologies such as InternetOfThings (IoT) sensors, Big-Data analysis, predictive and condition-based maintenance, intelligent logistics, intelligent asset management, etc. Hence, providing a state of automation for the whole chain of procedures and reliable data quality is of paramount importance.
\end{abstract}
\section{Background}

Infrastructure comprises the basic physical and organizational structures and facilities (e.g., buildings, roads, bridges, power supplies) needed for the operation of a society or enterprise. Infrastructure, by definition, is spread in vast geographical areas whenever the operation is required. Therefore, infrastructure is expensive to build and crucial to guarantee optimal operational conditions. 
For example, according to European statistical sources, the highest motorway density in Europe is found in the Netherlands (78 km per 1000 $km^2$ on average in 2009).  In the highly developed highway system of the Netherlands, the costs of infrastructure investment are high “despite the favorably flat landscape” because they “are inflated by high population density, legal and regulatory aspects (e.g., complicated and lengthy land-freeing procedures, environmental regulation), a high number of crossings with existing infrastructure and waterways and problems associated with building on wetland, particularly reclaimed from the sea.” According to a 2007 study, it takes over twenty years from the proposal of a road to its construction, which may reduce the incentive to handle congestion by infrastructure-based solutions. \cite{noauthor_how_2010}
Bridges are transport infrastructure components that require special maintenance attention. Structures such as bridges are characterized by large investments and 50 to 100 years long service life. Although the annual maintenance cost is relatively small compared to the investment cost (less than 1\%), the sum of the maintenance cost over the service life is of the same order of magnitude as the investment cost.\cite{klatter_9th_nodate}
Windfarms are another type of infrastructure spread in vast on and offshore areas that need constant maintenance to optimize the costs and downtime losses. There are 2,525 onshore wind turbines, generating a total of 3,000 MW of electricity, around 5\% of the Netherlands’ total requirement \cite{zaken_wind_2017}, and four offshore windfarms generating 957 MW \cite{noauthor_existing_nodate}.
The same can be said about other types of concrete infrastructures \cite{takewaka_maintenance_nodate}:
The basis of infrastructure maintenance is that the structure is adequately maintained such that its performance is always above the required level during its service life. However, since different materials are used in different structures such as buildings, dams, bridges, etc., which perform under different environmental conditions, it is not possible to lay down identical performance criteria for all structures. Therefore, the maintenance action is classified into four different categories specified as follows:

\begin{itemize}
    \item Preventive maintenance - the maintenance to prevent the appearance of visible deterioration on the structure during the service life.
    \item Corrective maintenance - the maintenance in which, appropriate countermeasures should be taken after degradation appearance of the structures has appeared.
    \item Observational maintenance - the maintenance carried out primarily on the basis of visual inspection without any direct measure and permits certain deterioration of the structure.
    \item Sensory inspection maintenance - the maintenance applied to the structure in which the direct inspection is difficult or practically impossible to be carried out, such as underground structures.
\end{itemize}
Critical structures such as dams and nuclear power plants having a long service life or structures situated in very harsh environments may be classified into the higher maintenance category. Similarly, criteria for classifying structures into other maintenance categories need to be developed. It should be pointed out that certain structures in which any maintenance action is very difficult to carry out may be categorized separately. 

This project aims to reveal that an infrastructure authority cannot avoid the need to introduce technologies to monitor the condition of its assets and to predict when assets will fail. It suggests that authorities must be proactive in searching for the best technologies for their purpose.
Research and experience have shown that maintenance is a core process in ensuring that infrastructure assets are optimally and functionally available to support business operations. However, the main challenge is the lack of skilled and experienced personnel to understand and anticipate maintenance requirements. A second challenge is the reduced time window available to carry out inspection and maintenance works. To overcome these challenges, authorities invest in technologies. However, technologies available to facilitate this process are complex and constantly shifting. Consequently, there is a need for infrastructure authorities to develop their technology absorption capabilities, i.e., the ability to embrace and capitalize on new technologies to enhance their maintenance management process. \cite{too_infrastructure_2012}

\subsection{Prognostics Health Monitoring}
In predictive maintenance, the challenge is to detect or foresee an upcoming failure in time, such that repair or replacement can be done before the system fails. This is often called condition-based maintenance (cbm), since maintenance is only performed when the system condition actually requires it. The basic way to achieve this is to monitor the system (either by continuous monitoring or by periodic inspections) to obtain a timely diagnosis of a degrading system.
Smart maintenance for Prognostics and Health Monitoring to Enable Predictive Maintenance is the new terminology emerging from the latest developments in the field of Internet-Of-Things, Big-Data analysis, and predictive and/or condition-based analytics. These components combined together can provide enough feedback to implement a new typology of maintenance focused on preventing the anomaly, which is associated with a contract penalty, to happen by predicting and addressing the failure moment \cite{lughofer_physical_2019}. 

\subsection{Process mining and Business Process Management}
The Business Process Management ﬁeld addresses the design, improvement, management, support, and execution of business processes. New data-centric process modeling approaches are a relevant and timely stream of process design approaches useful for the maintenance business process. \cite{reijers_evaluating_2017}
A Business Process Management System (BPMS) is an enterprise information system that ensures that a computer system can automatically allocate work to resources – humans and applications – in accordance with a predefined schema of the process, the available resources, and their dependencies. Over the years, it has become fashionable to refer to systems that work in this way as process-aware information systems or business process management suites. While commercial BPMSs have been around since the mid-1980s, the industrial appetite for this technology is still rising as more and more aspects of business processes are becoming subjects of constant data generation and collection.\cite{reijers_effectiveness_2016}
Most maintenance organizations have their versions of BPMS in the form of Maintenance Management Systems (pavement PMS, bridge BMS, windfarm WFMS), however, these systems are mostly indigenous, heterogeneous and human centric (relying in human input). However they provide precious knowledge about the field of interest. 

\subsection{Operation management}
Operations management is the activity of managing the resources that create and deliver services and products. The operations function is the part of the organization that is responsible for this activity. Every organization has an operations function because every organization creates some type of services and/or products. This holds true also for the Maintenance and Asset Management organizations involved with supervising and maintaining the state of the infrastructure \cite{slack_operations_2010}.

\begin{figure}
    \centering
    \includegraphics[width=\textwidth, keepaspectratio]{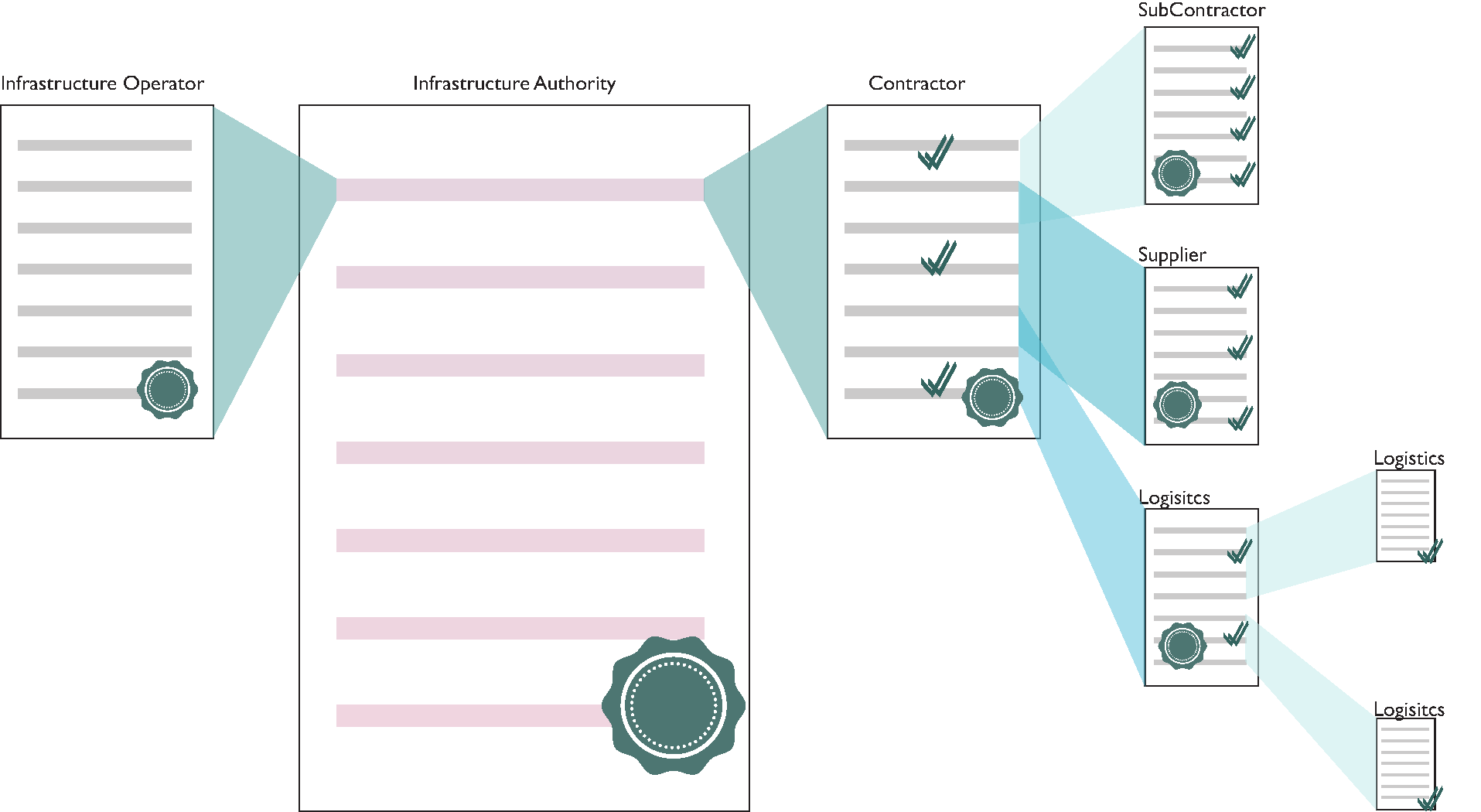}
    \caption{Traditional Infrastructure Maintenance Workflow}
    \label{fig:traditional_contract}
\end{figure}

Traditionally the maintnance procedure is compiled as a set of documents in the form of contracts and subcontracts as shown in Figure~\ref{fig:traditional_contract}. 

However, by combining these four paradigms, we propose a novel idea where the Infrastructure Maintenance is defined and stipulated into terms of a Smart Contract, terms that will guarantee the prompt maintenance reaction to the infrastructure degradation. 
\subsection{Cases}
\begin{figure}
    \centering
    \begin{minipage}{\textwidth}
    \includegraphics[width=\textwidth, keepaspectratio]{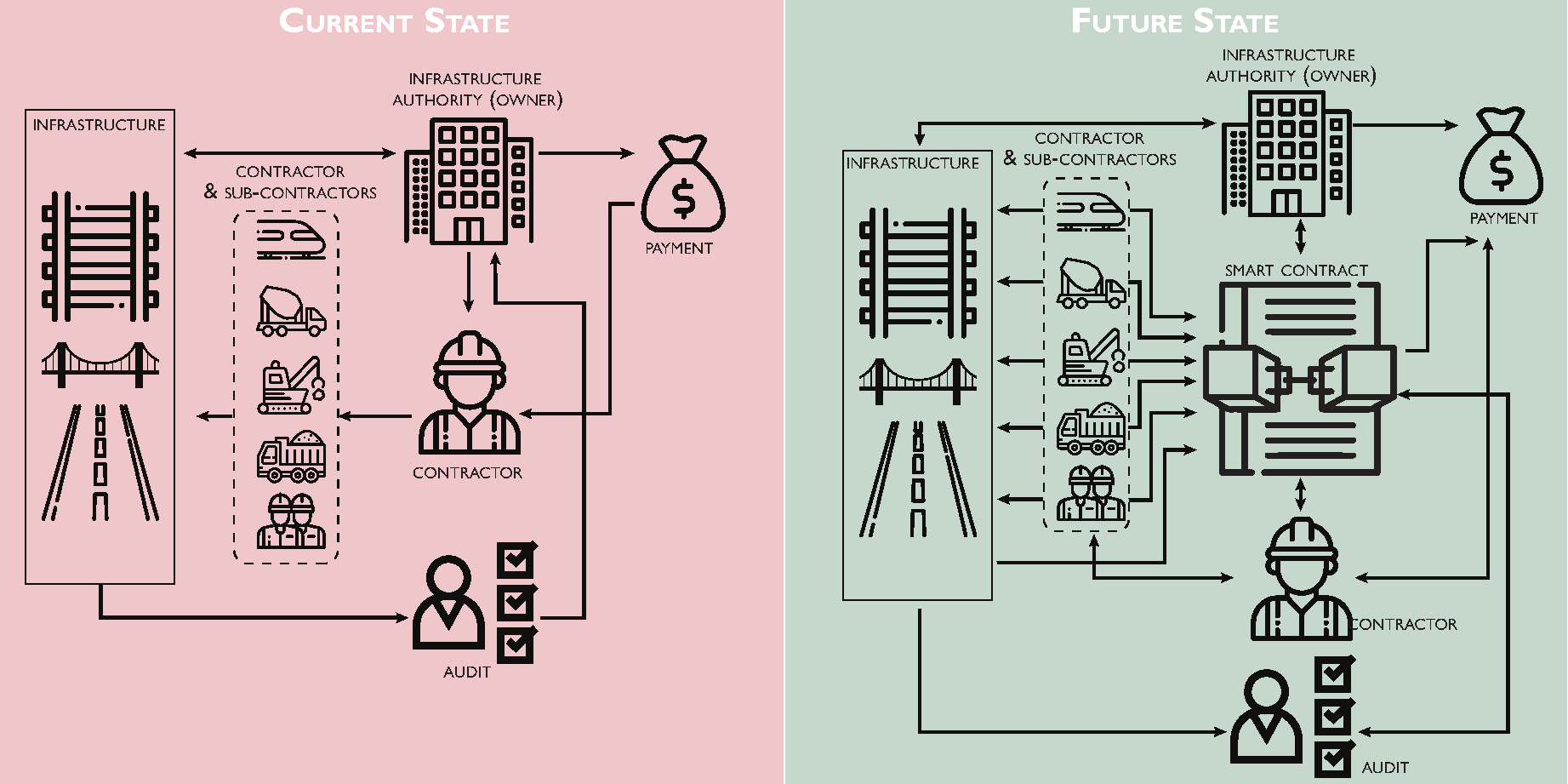}
    \caption{Transport Infrastructure Maintenance}
    \label{fig:smart_prospect}
\end{minipage}
\begin{minipage}{\textwidth}
    \centering
    \includegraphics[width=\textwidth, keepaspectratio]{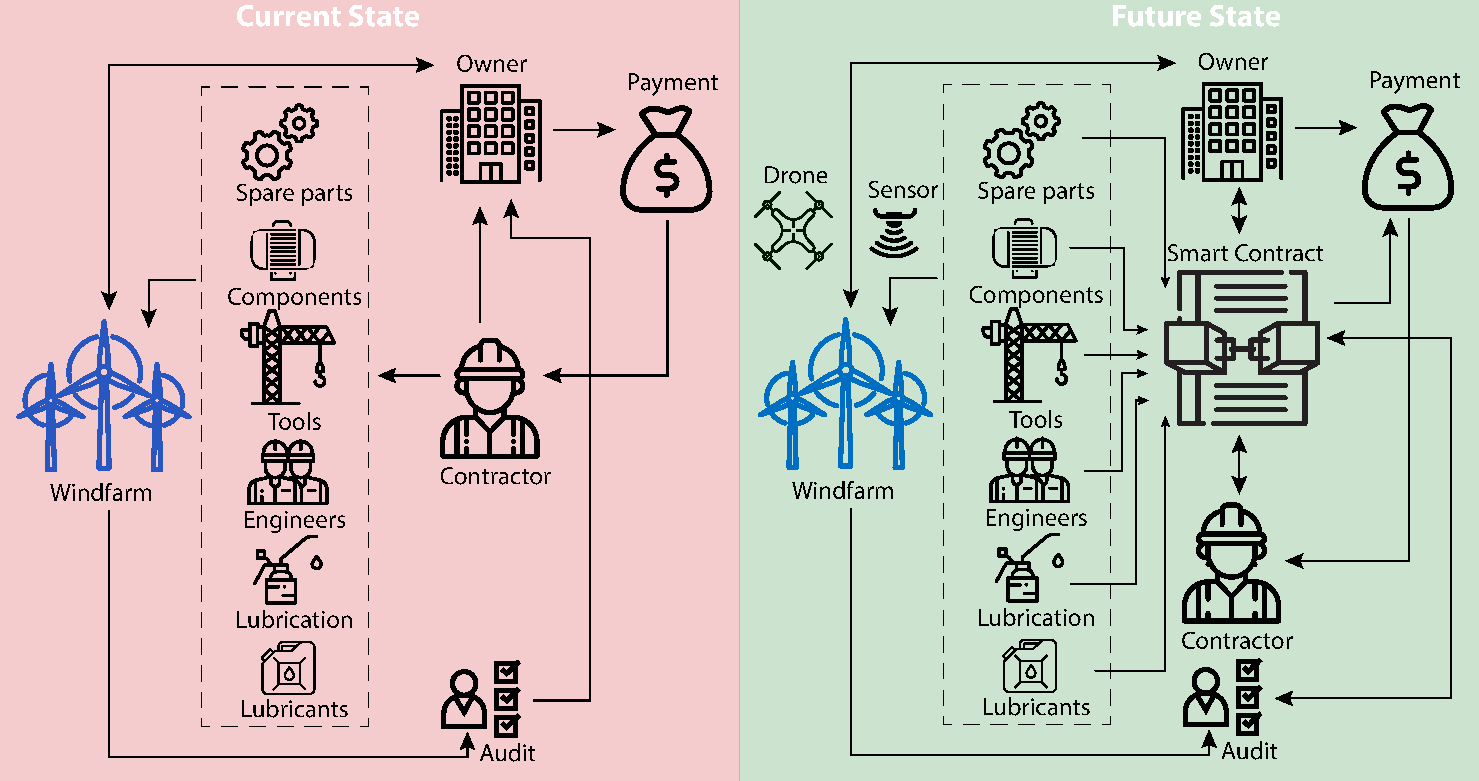}
    \caption{WindFarm Infrastructure Maintenance}
    \label{fig:smart_windfarm}
\end{minipage}
\end{figure}

Let us consider the case of infrastructure maintenance. The asset is owned by one party and maintained by a contractor  conforming to a strict contract policy. The contract stipulates severe penalties for all the downtime of the asset. On the other hand, to avoid the costly penalties, the contractor has to foresee all the replacement parts, the man labor, and other costs based on a given maintenance model.
From a smart contract perspective, if a smart monitoring system triggers all the defined rules foreseen in the contract, the process will become automatically more efficient and less costly. The delays introduced by the human factor will vanish, and liability will become more evident. For example, in the case of a continuous intelligent monitoring system, a concern is raised for a particular segment of the infrastructure. This event will trigger the contractor to take adequate measures to address the situation by initiating a transaction with the supplier of the materials required by the maintenance procedure, as well as allocating the assets necessary for the rehabilitation of the segment and allocating the time slot for the maintenance activity by the infrastructure operators. In this scenario, all the parties involved in the agreement described by the contract terms will automatically fulfill their share of liabilities. All the transactions will be executed automatically and approved by all the parties. 
Figures \ref{fig:smart_prospect}\ref{fig:smart_windfarm} show how the implementation of smart contracts transforms and facilitates the whole maintenance procedure in two different industries, transport infrastructure and windfarm infrastructure.

To achieve a smartcontract based maintenance, a reliable smart-monitoring system is of paramount importance.

\subsection{Challenges}
Although all infrastructures can be characterized and generalized with common structural and administrative features the fact remains that they are still very diverse. This heterogeneity makes the process of building a generic framework for infrastructure maintenance a challenging task. The challenges start from the design of sensing algorithms to the network protocol used for the data transmission in large scale. Designing an appropriate balanced data and model driven prognosis system. Learning all the process information involved with the maintenance process and designing an automatic response operational system based on distributed ledger topology in the form of a smart-contract.
Albeit a plethora of cheap useful sensors are becoming more common, the fact is that these sensors are not designed for highly accurate sensing.
An industrial high-end sensor should be highly accurate, reliable, and robust, whereas these characteristics are compromised in commercial IoT sensors. The quality of a sensor on such devices is low \cite{seraj_rolling_2017}. Using these sensors as sensor nodes for infrastructure monitoring confronts us with the following challenges:
\begin{itemize}
	\item Sensor quality \textemdash cheap MEMS sensors with declared accuracy settings.
	\item Sampling rate \textemdash Infrastructure monitoring requires sensor that measure low frequency components, which are bulky and expensive, commercial sensors have higher bandwidth operation.
	\item Localization \textemdash Geo-location estimation accuracy of the commercial GPS chip is in the order of 8 meter, which is larger than the state of the art measurement vehicles.
	\item Uncertainties \textemdash The position of the sensor withing the structure relative to the stress points.
	\item Networking \textemdash Because the infrastructure deteriorates slowly, starting from specific weak points and the data transmission hurts the energy consumption opportunistic wireless mechanisms should be designed from scratch.
	\item Cloud Computing \textemdash When such systems deploy in large scale, the destination data warehouse will overwhelm with information making a challenge to aggregate, visualize and process the information. Making participation of any sensor in the process of infrastructure monitoring easy, a consistent monitoring model is needed. In other words, the system should be available for a large-scale heterogeneous deployment.
\end{itemize}
\begin{figure}
    \centering
    \includegraphics[width=\textwidth, keepaspectratio]{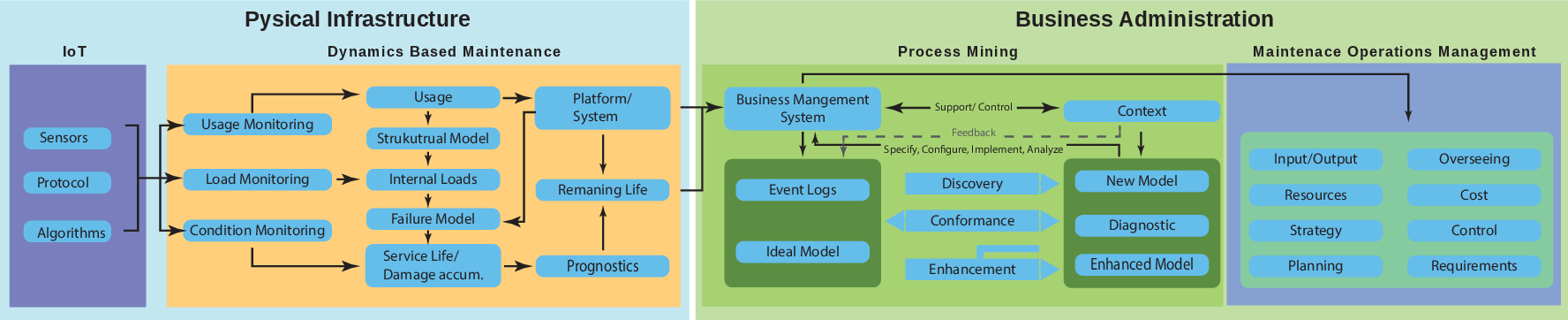}
    \caption{Framework flowchart representation}
    \label{fig:framework}
\end{figure}
\subsection{Research objectives and hypothesis}
The main focus of proposal is on design and evaluation of infrastructure maintenance systems that are cheap, robust, real-time, and easy to deploy also can act as alternative/complementary to the current maintenance technologies. 
A graphical representation is described in Figure \ref{fig:framework}.
To this end, the focus is to investigate the potential of reducing the influence of the human factor decision making in front of huge multidimensional information. We aim to provide a framework and associated services/functions that work consistently over all infrastructure types by studying in detail both physical infrastructure and business administration associated with it.

\subsection{Research questions}
The main research question to be answered is: 
\begin{quote}\itshape{How to create an ecosystem by integrating all the latest information technological advancements for a seamless infrastructural maintenance??}
\end{quote}

In order to answer this question and address the aforementioned challenges, the following encapsulated sub-questions need to be answered first:
\begin{quote}
\begin{itemize}\itshape{
\item What is the nature of infrastructure anomalies, which sensors capture them best and how the captured information should be presented to the aggregation/modell point?
\item How to build a generic and expandable prognostic model framework to satisfy different types of maintenance and asset management industry?
\item How to build an autonomous knowledge base system by \textit{mining} the knowledge from the existing traditional Maintenance Management Systems and combine with the aforementioned prognostic model framework?
\item What are the operational steps required to execute a maintenance procedure and update the knowledge base with gained feedback?
\item How can all these steps be combined into a set of automatic executable rules in the form of a smartcontract? 
}
\end{itemize}
\end{quote}

\subsubsection{Research hypotheses}
The research starts from the following hypothesis:

\begin{itemize}\itshape{
    \item \textbf{Hypothesis 1}. Despite the fact that commercial sensors lack the accuracy of state-of-the-art ones, in the context of IoT systems, they will be deployed in huge scales and provide continuous information for decades, thus compensating the uncertainties introduced by measurement accuracy with the volume of data.
    \item \textbf{Hypothesis 2}. Using state-of-the-art signal processing, machine learning, cloud computing, and advances in predictive maintenance, the collected infrastructure sensor data can be classified into structural health states as described by the standartized requirements.
    \item \textbf{Hypothesis 3}. Existing Maintenance and Asset Management Systems have accumulated a huge amount of knowledge, thus using state-of-the-art process mining technologies to extract and investigate that information will allow us to find common features and functionalities among different infrastructure typologies.
    \item \textbf{Hypothesis 4}. The maintenance process involves a set of common operations that span from provision of raw materials, logistics, staff rostering, repair work etc.. Operations that can be generalized and used as adaptive templates among industries.
    \item \textbf{Hypothesis 5}. While traditionally all the steps, rules and liabilities of the maintenance procedure are sealed in ink written paper contracts, the same steps, rules and liabilities can be described as computer code sealed in a smartcontract.}
\end{itemize}

Building upon the aforementioned hypotheses, our approach starts with identifying the targets, challenges, and solutions.
Being a complex field, dominated by decades long research and field experience with sophisticated equipment, the maintenance indicators are well identified and classified for each infrastructure type.

Because infrastructure maintenance affects a wide range of industries, the interest and incentive to adapt such technologies, methodologies and framework is obvious.
\section{Smart Contract}
SmartContracts are first mentioned by Nick Szabo in his 1996 paper \cite{szabo_smart_1996} he created the notion of a decentralized ledger that could be used as tool to agree the automatic completion of the stipulated parts of the contract. This type of digital contracts are knows as  self-executing contracts or block-chain contracts.
Szabo \cite{szabo_secure_2014} initially proposed that smart contract infrastructure can be implemented by replicated registries and contract execution using cryptography hash chains and Byzantine fault tolerant replication. In the meantime the register will be updated by the feedback of the transaction services.
Benefits:

Because the smart contract is stored as an executable code in a encrypted peer to peer network, everything that gets recorded in the blockchain becomes immutable and cannot be changed, with copies distributed among all the parties involved in the contract, the smart contract provides benefits that outperform the efficiency of traditional contract when applied in certain domains.
Such benefits are:
\begin{enumerate}
    \item Transparency. The rules and conditions stipulated by the contract are transparent to all the parties and cannot be changed unilaterally without the agreement of all the parties. Another aspect of transparency, is the transaction transparency between the involved parties.
    \item Accuracy. Running in a procedural code, where the procedure involving parties is explicitly described, the contract will only execute when the conditions is met. Avoiding the errors introduced by the traditional paperwork procedure.
    \item Speed. Speed is achieved because contracts execute automatically in the network, enhancing efficiency. Thus, reducing the communication time to zero.
\end{enumerate}

The security of the contract is satisfied by the characteristics of the blockchain. Namely the immutability. This feature provides the following benefits:
\begin{enumerate}
    \item Security. The block-chain architecture stores the information in encrypted hash records, distributed and proved by all participants and.
    \item Traceability. Any of the terms of the contracts, once executed in a transaction is automatically stored as hash block in the chain. They can be traced through all the historic chain of product development or service provided, helping for a rapid response to contingencies.
    \item Trust. The nature of the code upon which the smart contract is written provides the confidence on the execution of the contract. Providing a safe, transparent and autonomous execution, diminishes the possibilities of manipulations or errors. 
\end{enumerate}

\begin{figure}
    \centering
    \includegraphics[width=\textwidth, keepaspectratio]{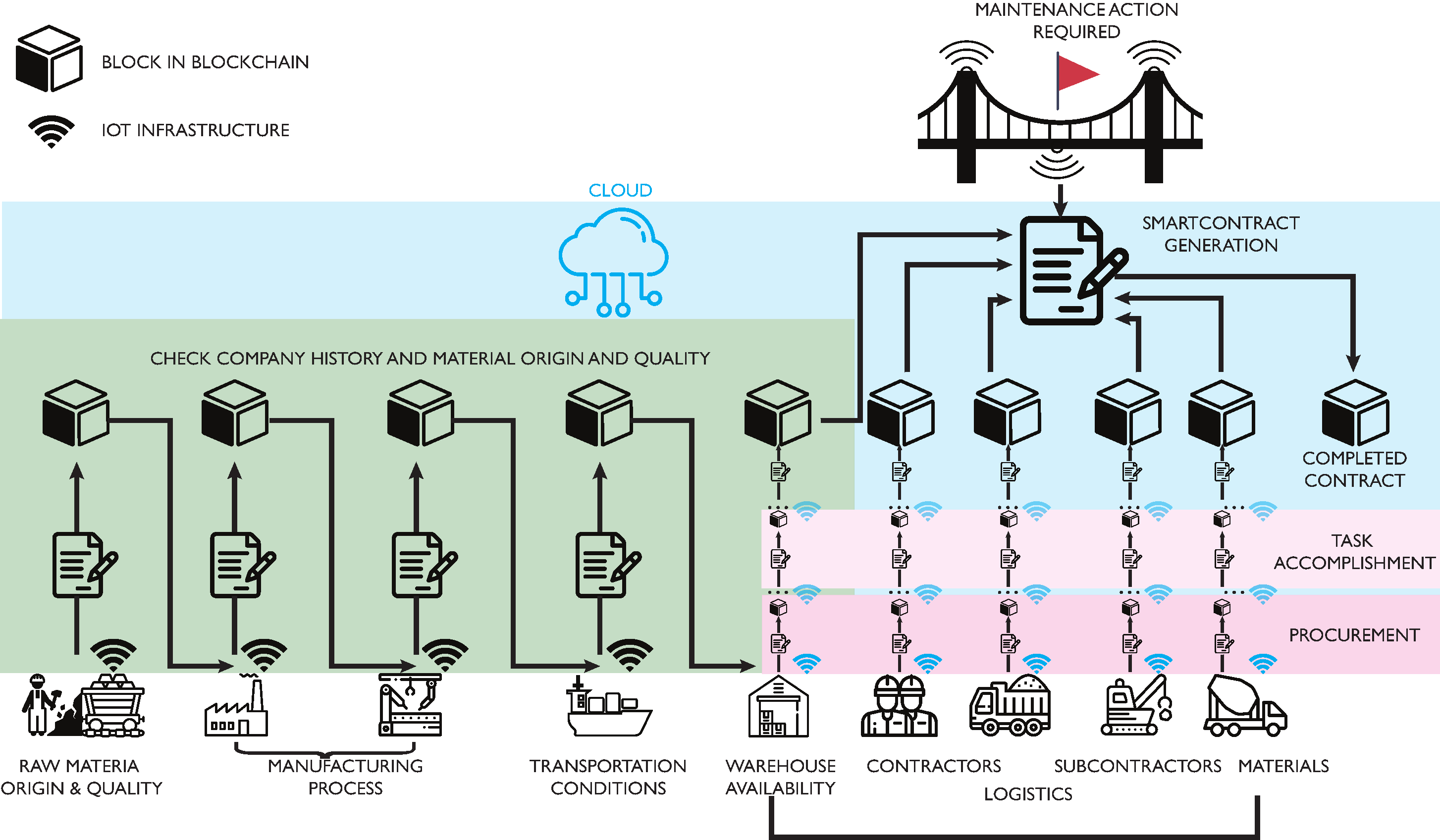}
    \caption{The whole operational bridge maintenance procedure expressed as a smartcontract}
    \label{fig:smart_contract}
\end{figure}
\section{Internet of Things IoT}
\subsection{Interpretation of Monitoring Data}

In the present era sensors are ubiquitous, all systems are connected (internet of things, IoT) and data storage is not an issue anymore. In practice many original equipment manufacturers (OEM) apply a lot of sensors to the assets and systems they produce, enabling the owners and operators to collect a lot of data on their systems, with the promise that the system can be maintained condition-based. However, just applying a number of sensors to a system does not mean that the condition of the system is assessed. Translating the collected raw data into useful information on asset condition is in many cases challenging. Sometimes just observing a trend in a monitored parameter, or comparing the measured value with a predefined threshold provides the required insights. But in most cases this is not sufficient, and a thorough understanding of the normal/dynamic system behavior, as well as the system failure behavior is required. For example the vibration monitoring of bearings is a field that is so well developed (after being in use for many decades), that understanding the details of bearing (failure) behavior is not needed to properly diagnose a faulty bearing. But interpreting the vibration behavior of a more complex system like a bridge or wind turbine rotor blade, aiming to detect damage, is much more challenging.

\subsection{Data} 

The key to the successful implementation of predictive maintenance methods are data collection and processing. Terminologies such as \textit{smart systems}, ‘smart industries’ and other combinations with the word ‘smart’ all refer to the usage of sensors in a product, system or installation. These sensors provide the necessary data to assess the current loading and/or performance of the structure being monitored. However, this data collection does not yet transform the system into a smart system. 
The smartness is embedded in the processing of the data, converting it to information, and the subsequent decision process. It is clear though, that decisions cannot be made in the absence of data: data are a prerequisite. The question is whether it is possible to define upfront what data is needed, which sensors are best suited for this task and which signal processing techniques are to be applied. This is not a sequential design process, but more a parallel and iterative process: on the one hand, the selection of signal processing techniques influences the choice of sensors and hence determines which data can be made available. On the other hand, the system being monitored and the conditions in which the monitoring should take place set constraints on the sensors and thus dictate the choice of sensors. These inherently delimit the choice of processing techniques from a different angle.

\subsection{Sensors}

The first step after having established the importance of the acquisition of data is creating a categorization of suitable sensor technologies and data acquisition systems. We focus on sensors and systems suitable for dynamic or vibration measurements. One of the most frequently selected sensors is the strain gauge. Its ease of application and broad experience of application are the most common motivations for selecting this type of sensors. A strain gauge is a passive sensor in the sense it can only sense. Strain gauges can measure both static and dynamic strains and thus provide data on the local strain field, hence the use of strain gauges aligns well with the concept of load monitoring. Typical applications are fatigue dominated structures, in which a link between dynamic loads and the consumed fatigue life is established. 
Accelerometers are widely used to capture the dynamic response of structures. Accelerometers are passive sensors and in most cases need an external power supply to function. Accelerometers operate in a specific frequency range. Frequencies outside this range, bound by a higher as well as a lower limit, are not captured accurately. The lower the lower bound of the frequency range, the more bulky the accelerometers gets. MEMS-based accelerometers, such as those found in mobile phones, have pushed the use of accelerometers, bearing in mind that their size limits the lowest frequency that can be measured. However, the level of integration that can be reached with these devices and their significantly more favorable energy consumption are strong advantages. The lack of signal quality can be compensated by following a crowd sensing big data approach. 
Piezo-electric transducers (PZT) have excellent options for integration into the structure at a relatively low cost. Moreover,PZTs can be used both in sensor and actuator mode, which makes them very flexible in use. The frequency range in which PZTs can be used is also very broad, be it that excitation at lower frequencies typically requires more power than the (average) PZT can produce. PZTs are applied to measure the structural dynamic response or the nonlinear response of, e.g. composite structures, as well as to generate and measure propagating waves, such as guided waves in composite materials or ultrasonic waves in, for example plastics and cement materials. In sensing mode, piezo-electric transducers are passive sensors: no power needs to be supplied. The mechanical motion of the structure causes a current to flow as a direct result of the piezo-electric effect. This opens the door for another application of PZTs: energy harvesting. Energy harvesting is a key element for smart and autonomous sensor nodes, as they either rely on batteries having a finite, relatively short endurance or local energy generation.
Humidity sensors (Hygrometer) senses, measures and reports both moisture and temperature. The ratio of moisture to the highest amount of moisture at a particular temperature is called relative humidity. Relative humidity becomes an important factor, when measuring the moisture around or inside the infrastructure. Humidity sensors come in three types: Capacitive, Resistive and Thermal, with their own typical applications. Some parameters to judge the appropriate humidity sensor for infrastructure scenario are the accuracy, reliability, response time and the energy consumpation.

\subsection{Wireless Sensor Networks}\cite{ayele_towards_2018}

The past few years, have witnessed the steady rise in utilization of Internet of Things (IoT) technologies. Infrastructure monitoring is one of the trending applications of IoT technologies, where a number of heterogeneous sensors are deployed to monitor the state of the infrastructure in geographically vast areas. These sensors could be deployed inside the infrastructure or around it, to monitor the structural changes and deterioration. The sensors will operate in with a limited source of energy. To these ends, it is important to achieve high energy efficiency, good reliability and low latency for a responsive maintenance strategy. Because the infrastructure is static by nature the deployment of these sensors facilitates some local data pre-processing and sharing among end-nodes for collaborative decision making. The rationale behind this is that data processing consumes less energy than transmitting large amount of data. Thus, instead of reporting all the raw data generated from the end-devices, some sort of local decision making from neighboring nodes allows the relay of the aggregated data at a reduced energy consumption to the central server. Wireless sensor networks with conventional routing algorithms (e.g DSR or AODV) can be used, but they often perform poorly in scenarios where the communication path is intermittent due to the location of the infrastructure to be monitored. Opportunistic mobile networks (OMNs) are the recent evolution of the traditional mobile wireless sensor networks (MWSN), which are used to provide communication facilities among devices in sparse and mobile network scenarios, like infrastructure monitoring applications. Several opportunistic data dissemination schemes, on sensor data gathering, have been proposed in several recent studies, e.g. Epidemic, Spray and Wait (SnW), Direct Contact (DC), etc. These networks are developed on top of IEEE802.15.4 standard, which often suffers from huge overheads due to the complex implementation. More recently, the energy efficient version of Bluetooth known as BLE - Mesh, has better energy efficiency and wider coverage over IEEE802.15.4 based solutions. While making some progress in energy efficiency aspect, these solutions still suffer from high latency mainly due to node mobility. Although this issue can be addressed by using complementary long-range wireless technologies (e.g., cellular) to facilitate connectivity, they often consume higher power. Fortunately, emerging low power wide area network (LPWAN) technologies such as LoRa, Sigfox or NB-IOT promise to provide better coverage with a low energy consumption that seem to support many requirements of infrastructure monitoring application. LPWANs in general are fundamentally designed to ensure very long battery lifetime and provide seamless interoperability among end devices without the need for complex local installations. Among LPWAN technologies, LoRa is believed to have high potential for realization of LPWAN IoT goals. However, IoT applications requiring high data rate and low latency are not particularly the strength of LPWANs mainly due to the generic low bitt-rate, stricter duty-cycle restrictions and the larger packet header associated with it. Thus to realize the WMS design requirements, a mechanism to control the trade-off between energy versus latency is necessary, which is not practically achievable by using only LPWAN technology. A favorable approach would be to  integrate OMN and LPWAN infrastructure.
\section{Physical Model-Based Prognostics and Health Monitoring to Enable Predictive Maintenance \cite{lughofer_physical_2019}}

Nowadays, only a limited number of systems are operated in completely stable conditions. Most of the systems, like wind farm turbines, infrastructures are facing largely variable operating conditions and environments. At the same time, failures in any of the associated subsystems or components may have large consequences, e.g. high costs (loss of revenues, high logistics costs due to remote locations) or large safety and environmental impacts. To control the number of failures, typically preventive maintenance at predetermined intervals is performed. By replacing the components in time, failures can be prevented, but this is a rather expensive policy when the operational proﬁle is largely varying. The preventive maintenance intervals must be set to very conservative values to assure that also severely loaded subsystems do not fail. This is a costly process, but it also limits the availability of the system, as it must be available for maintenance tasks quite often. To improve this process, reduce the costs and at the same time increase the system availability, a better prediction of failures for systems operated under speciﬁc (and mostly dynamic) conditions is required. Only when such a prediction is available, maintenance can be performed in a just-in-time manner. This is the promise that predictive maintenance as the ultimate maintenance policy is giving. However, although a lot of research has already been done on this topic in the past decade, still a generically applicable concept is not yet available. This chapter will discuss the challenges encountered in developing predictive maintenance concepts, and will provide insights and decision support tools that can assist in further improving the existing methods.

Monitoring of Usage, Loads, Condition or Health
To be able to diagnose a system, structural health monitoring (SHM) or condition monitoring (CM) systems can be deployed. Although the purpose of both SHM and CM is to assess the condition of a system, their origin and approach are slightly different. 
\begin{enumerate}
\item SHM methods are typically focusing on measuring and interpreting the dynamic response of a system, aiming to detect, localize and quantify damage. 
\item CM covers a wide range of techniques, measuring various quantities that can indicate an upcoming failure in the monitored system. For rotating equipment, vibration monitoring is a well-known CM technique, but also lubrication oil analysis and corrosion monitoring can be considered as CM.
\end{enumerate}

One of the challenges associated with monitoring is to decide for each speciﬁc application what the most suitable monitoring strategy is (usage, load, condition), what quantity should be measured at which location and what sensor type is preferred.

Data-Driven or Model-Based Prognostics
Prognostics can be based on either data-driven or model-based approaches. The data-driven approaches use large amounts of data, preferably from various sources, and apply data analytic techniques like machine learning and artiﬁcial neural networks to discover patterns and relations in the data sets. This means that in principle no knowledge on the system characteristics or failure behaviour is required, which makes the approach popular and widely accessible. However, the lack of system knowledge can also lead to the discovery of trivial or accidental(non-casual) relations. For example, a high correlation between vehicle ﬂow and the pavement wear could be discovered from a data set, but that relation is trivial from an engineering point of view. Further, the artiﬁcial intelligence (AI) methods used in this approach must be trained with data to enable to learn the patterns. This means that in principle only patterns(i.e.failures) can be predicted that have been observed before (and were included in the training set). The alternative approach for prognostics is the use of physical failure models. In this approach the failure mechanism, like fatigue, wear or corrosion, is captured in a mathematical model, relating the usage or loading of a system or component to a degradation rate or lifetime prediction. Monitoring of the usage or loads on an individual system then enables to predict the (remaining) time to failure. Although the development of these types of models is rather time-consuming, they solve some of the limitations of the data-driven approaches: the models do not need a large set of failure data (which is typically not available for critical systems), and they also work for situations not previously encountered. The main challenge on this topic is to decide which approach is most suitable in a speciﬁc situation.

\section{Business Process Management}

Process Mining  Process mining is a family of techniques in the field of process management that support the analysis of business processes based on event logs. During process mining, specialized data mining algorithms are applied to event log data in order to identify trends, patterns and details contained in event logs recorded by an information system. 
Process mining aims to improve process efficiency and understanding of processes.
Process mining is also known as Automated Business Process Discovery (ABPD).However, in academic literature the term Automated Business Process Discovery is used in a narrower sense to refer specifically to techniques that take as input an event log and produce as output a business process model. 
The term Process Mining is used in a broader setting to refer not only to techniques for discovering process models, but also techniques for business process conformance and performance analysis based on event logs.

The Business Process Management (BPM) ﬁeld addresses the design, improvement, management, support, and execution of business processes. Its cornerstone artifact is the process model, which is used to describe the activities to be executed to handle a case, as well as their order of execution. There is a range of modeling approaches available, with two main categories of these being the more traditional, activity-oriented process models and the more recent, data-centric approaches. 
In the last decade, a number of data-centric approaches have been developed to extend or counter the overly activity oriented way of designing and executing business processes, as activity-oriented process models had become known for a number of limitations. For example, in dynamic environments, they can become too rigid, which makes it difﬁcult to cope with changes in the process. The central idea behind data-centric approaches is that data objects/elements/artifacts 
can be used to enhance a process oriented design or even to serve as the fundament for such a design. This has certain advantages, varying from increasing ﬂexibility in process execution and improving re-usability to actually being able to capture processes where data play a relevant role. Despite considerable research efforts, it seems fair to say that data-centric process approaches have not become mainstream at this point. Yet, the recurrence of new variants that treat data as a ﬁrst-class citizen suggests that the core idea in itself has merit. Therefore,we consider data-centric process modeling approaches as a relevant and timely stream of process design approaches to further examine with regard to the level to which they live up to their promises.

Despite relentless information technological advances, organizations struggle to create value and improve their business processes. Process models turn out to be a helpful tool to improve the understanding of current business operations and are often used as a foundation for improvement initiatives. Yet, any organization that engages in modeling its business processes, inevitably encounters questions. 
Which processes exist in my organization? 
Where does one process end and the other begin? 
At what level of detail should these processes be modeled? 
These questions are especially relevant when an organization’s interest in its processes results in a collection of hundreds of process models. 
The notion of a business process architecture can be used to address questions like the ones we mentioned.
Such an organized overview of the processes that exist within an organizational context, along with the guidelines on how the related models should be organized, is what can help individual modelers to arrive at a consistent and integrated collection of process models.

However, the introduction of a business process architecture clearly begs the question of how in any given situation it should be established. Given the variety of views on how to design a business process architecture, we identify a lack of understanding of the diﬀerences between these views and uncertainty among business users to make the right choices. Yet, it has been recognized that not any business process architecture is equally eﬀective. In a recent blog post, for example, Derek Miers notes how some process architectures may actually impede on the process-centered line of thinking that was behind the initiative to model process in the ﬁrst place. We have to investigate which approaches and guidelines to create a business process architecture, which of these are considered most useful in and how they are actually applied in practice. In pursuing this aim, we should focus on the conceptual, modeling-related, challenges involved in organizing a business process architecture, rather than the organizational or socio-political issues that play a role in the eﬀective use of process models. More precisely:
\begin{enumerate}
    \item The design approaches and guidelines that can be used to design a business process architecture. 
    \item Provide an exploratory comparison of the use and usefulness of these design approaches and guidelines in practice. 
    \item Organize a framework that is aggregated from existing approaches and that can be used as a basis for structuring a maintenance process architecture.
\end{enumerate}

To arrive at these contributions, ﬁrst, a structured research of the existing maintenance processes has to be conducted. 
Leading to an overview of the design approaches, as well as an overview of the guidelines and structuring principles that these approaches propose to create a process architecture. 
Next, the use and usefulness of the diﬀerent design approaches in practice has to be explored. 
Based on this analysis the most commonly used structuring principles can be integrated in the form of a framework, to be used as a basis for structuring a new generic maintenance process architecture.
\section{Operations research and maintenance \cite{vieira_operations_2016}}
Operations research (OR) is a discipline that combines knowledge from fields such as applied mathematics, computer science, and systems engineering. It encompasses a wide range of techniques for improved decision-making, commonly for real-world problems. Originally, OR emerged as a way to improve military material production during the second world war but methods have continuously grown to model and solve problems in business and industry eversince.

During the last decades, a wide range of problems have been addressed to support strategic decision making, facilitate day-to-day maintenance management, and solve management problems related to the maintenance practice. Among the existing OR applications for maintenance management and logistics optimization, well-known problems include scheduling, staff rostering, resource planning and scheduling. Given the growing acceptance of OR models to solve problems in maintenance, research on modeling emerging problems receives increased attention. Both a taxonomy for resource capacity planning, control decisions and algorithms to solve the most relevant ones have been researched.

In a nutshell Operational Management objectives can be described as Plan, Track and Organize

Maintenance Operations Management (MOM) is mostly administered through Maintenance Management Software (MMS) systems.
The aim is to achieve optimum management of the capital-intensive assets of an organization. Three central components are assets, processes and logistics.
\begin{enumerate}

    \item Assets
The assets include all assets that are used to keep the business running: machines, it equipment, buildings, etc. Maintenance Management Software record the existence of the assets, follow them from the engineering phase, through the purchase and commissioning up to and including taking out of the service maintenance. The assets are managed during this period, so that all planned or unplanned work and changes related to the operation of the asset or the role within the organization are recorded.

    \item Processes
The processes are based on the services that an organization provides. A maintenance management organization has internal customers as the most important service customer. All processes surrounding serving this customer are efficiently modeled and managed through MMS. Service-oriented organizations will want to see the handling of questions, malfunctions, changes, etc. in one integrated maintenance management software system.

    \item Logistics
Organizations are increasingly dependent on the infrastructure to deliver their services to external customers. This means that infrastructure segments who experience disruptions should be quickly and effectively addressed. This can be done by implementing Incident registration in the maintenance management software, but also self service is a factor that is becoming increasingly important within the software.

\end{enumerate}
The maintenance chain of operations
The maintenance chain is characterized by a sequence of operations, which depends on the characteristics of the infrastructure (such as location, level of deterioration, etc.). Figure 1 depicts a deployment flowchart of the asset life-cycle and maintenance operation procedure. After the asset is constructed it becomes a subject of continuous health monitoring. The inspection is carried out to check for deterioration and to check the state after the maintenance work is carried out.
The operations that take place once the deterioration is detected during the routine inspections.
Inspections are scheduled based on the sensitivity of the infrastructure, usage or degradation level.
Once the degradation is identified the procedure for evaluating the severity of the degradation is planned.
This triggers the action to develop the necessary budget needed for the maintenance, deciding for the working force and equippments needed, logistics and repair plan.
The next step involves the repair work according to the plan. The repair work is checked to assess the effectiveness of the maintenance.
The system is updated with the feedback from the repair work, thus, updating the monitoring and maintenance plan for the repaired asset.

\begin{figure}
    \centering
    \includegraphics[width=\textwidth, keepaspectratio]{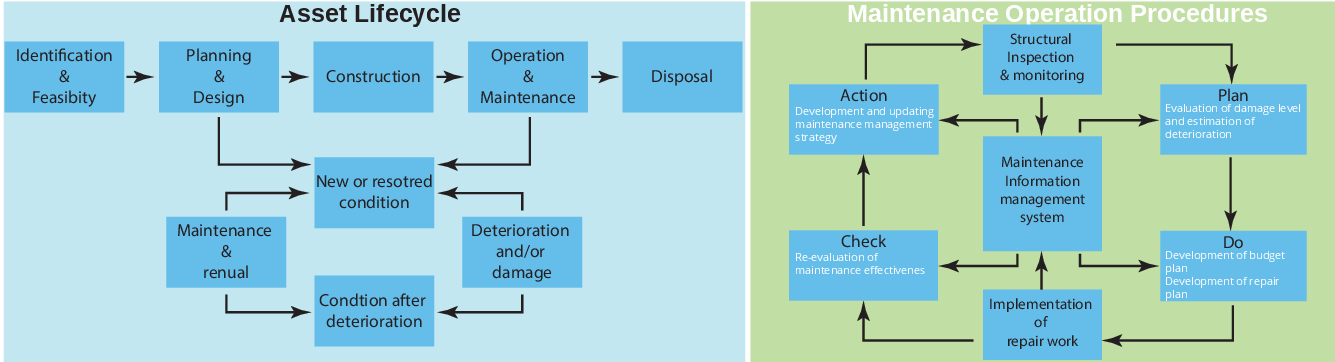}
    \caption{Asset lifecycle and Maintenance Operation Procedure}
    \label{fig:my_label}
\end{figure}

\bibliographystyle{abbrv}
\bibliography{references}

\end{document}